\documentclass{aastex62}

\usepackage{amsmath}
\usepackage{color}
\usepackage{booktabs}
\usepackage{multirow}
\newcommand{\tabincell}[2]{\begin{tabular}{@{}#1@{}}#2\end{tabular}}
\accepted{by ApJ}
\shorttitle{Growth of massive black holes at high-z}
\shortauthors{Li \& Cao}
\begin{document}
\title{Growth of massive black holes at high-z via accretion predominantly driven by magnetic outflows}

\correspondingauthor{Xinwu Cao}
\email{E-mail: xwcao@zju.edu.cn}

\author{Jiawen Li }
\affiliation{Shanghai Astronomical Observatory,	Chinese Academy of Sciences, 80 Nandan Road, Shanghai, 200030, China}
\affiliation{University of Chinese Academy of Sciences, 19A Yuquan Road, 100049,
	Beijing, China}

\author{Xinwu Cao}
\affiliation{Zhejiang Institute of Modern Physics, Department of Physics, Zhejiang University,
38 Zheda Road, Hangzhou 310027, China, Email: xwcao@zju.edu.cn}
\affiliation{Shanghai Astronomical Observatory, Chinese Academy of Sciences, 80 Nandan Road, Shanghai, 200030, China}
\affiliation{University of Chinese Academy of Sciences, 19A Yuquan Road, 100049,
	Beijing, China}

\begin{abstract}
Luminous quasars powered by accreting supermassive black holes (SMBHs) have been found in the early Universe at $z \ga 7.5$,  which set a strong constraint on both the seed black hole mass and the rapid growth of the SMBHs. In this work, we explore how the SMBHs are grown through Eddington limited accretion driven predominantly by magnetic outflows. Most angular momentum and the released gravitational energy in the disk can be removed by magnetic outflows, and therefore the mass accretion rate of the black hole (BH) can be high even if the disk is radiating at sub-Eddington luminosity. It is found that the SMBH with several billion solar masses discovered at $z\ga 7$ may probably be grown through chaotic accretion predominantly driven by magnetic outflows from a stellar mass BH, when the disks are radiating at moderate luminosity ($\sim 0.5$ Eddington luminosity) with mild outflows. We find that most SMBHs are spinning at moderate values of spin parameter $a_*$, which implies only a small fraction of quasars may have radio jets.
\end{abstract}



\section{Introduction} \label{sec:intro}
It is believed that supermassive black holes (SMBHs) grow mainly by swallowing the material through a gaseous disk around the hole \citep*[e.g.,][]{1969Natur.223..690L,1982MNRAS.200..115S,2002MNRAS.335..965Y,2004MNRAS.351..169M,2009ApJ...690...20S}. If the SMBHs accretion activity is turned on, strong radiation is produced by the luminous accretion disks, and they appear as active galaxies. Some of them have relativistic jets or/and outflows \citep*[e.g.,][]{{1988ApJ...328..114P,2010A&A...521A..57T}}. Quasars at high redshifts have been extensively studied \citep*[e.g.][]{1999ApJ...526L..57F,2003AJ....125.1649F,2014AJ....148...14B,2015Natur.518..512W,2016ApJ...833..222J,2016ApJ...819...24W,2017ApJ...839...27W,2018Natur.553..473B}, the highest redshift of which discovered so far is $z_{\rm Q}=7.54$ \citep[ULAS J1342+0928;][]{2018Natur.553..473B}, with black hole (BH) mass $ \sim 8\times 10^8 M_\odot$. In the concordance $\Lambda \rm CDM$ cosmological model, the age of the universe today is $ t_0 \approx 13.79 \; \rm Gyr $ based on the Planck $ \rm CMB $ observations \citep[][]{2016A&A...594A..13P}, while the age at the redshift $ z_{\rm Q}=7.54$ is $t_{\rm Q}\approx 0.69 \; \rm Gyr $.  Therefore, the SMBH in this quasar should have grown from a seed BH within less than $ 0.69 \; \rm Gyr $. It sets strict constraints on the mechanisms of the seed BH formation and rapid growth of the SMBH in the early universe, which are still open issues  \citep*[e.g.,][]{2004A&A...420L..23K,2005ApJ...620...59S,2006MNRAS.373L..90K,2006ApJ...650..669V,2009ApJ...696.1798T,2016MNRAS.459.3738I,2018MNRAS.476..366B}.

In order to reproduce such rapid SMBH growth through gas accretion within about half a billion years, one has to assume almost continuous Eddington limited accretion started from a heavy seed BH with $ M_{\rm seed} \gtrsim 10^5 M_\odot $ at redshift $ z\sim 20-30 $, though the formation process of such a heavy seed BH is another important issue \citep[\citealt{2008ApJ...686..801O,2016ApJ...818..184P,2017ApJ...850L..42P,2018ApJ...864L...6P}, and see][for a review on heavy seed BH  formation]{2010A&ARv..18..279V}. The calculations of \cite{2005ApJ...620...59S} show that a non-rotating BH will be spun up to an extreme Kerr BH while its mass is increased several times through accretion in a thin disk. The radiation efficiency of a thin disk surrounding a Kerr BH is substantially higher than $\sim 0.1$ for a non-rotating BH, which implies that the mass accretion rate is reduced for an Eddington limited accretion disk surrounding a Kerr BH. If sufficient gas is fed onto the BH, the mass accretion rate can be very high, and the disk is no longer geometrically thin. In this case, the luminosity of the disk can be super-Eddington and the radiation efficiency becomes lower than $0.1$ due to advection of energy in the disk, which is well described by the so-called slim disk model \citep*[][]{1988ApJ...332..646A}. Thus, an alternative way for rapid BH growth is to assume that the BH has experienced super-Eddington accretion phase with a relatively low radiation efficiency (i.e., slim disk phase) \citep*[][]{2004A&A...420L..23K, 2004MNRAS.352.1390M, 2013ASSL..396..293H,2014ApJ...784L..38M}. However, numerical simulations show that most of the gas is driven into outflows by the radiation force of the disk with super-Eddington luminosity, and only a small fraction of the supplied gas can ultimately reach the BH horizon \citep[e.g.,][]{1999MNRAS.310.1002S,2002ApJ...573..738H,2003ApJ...592.1042I,2014ApJ...780...79Y}. The analytic model of a slim disk with outflows indicates that the mass accreted by the BH is always limited to $\sim\dot{M}_{\rm Edd}$ ($\dot{M}_{\rm Edd}=L_{\rm Edd}/0.1c^2$) due to the mass loss in the outflows, while the luminosity of the disk is super-Eddington  \citep*[][]{2015MNRAS.448.3514C}. It implies that the growth of SMBH through a slim disk with outflows is not faster than that through normal Eddington-limited accretion.

The growth of BH either through Eddington limited accretion or super-Eddington accretion will inevitably leads to most BHs being rapidly spinning Kerr BHs in the Universe, which implies that the average radiation efficiency of the SMBHs should be much higher than $0.1$. However, this is inconsistent with the radiation efficiency $\sim 0.1$ constrained by cosmological evolution of SMBHs calculated with observed active galactic nuclei (AGN) luminosity functions \citep*[e.g.,][]{2002MNRAS.335..965Y,2009ApJ...690...20S,2010ApJ...725..388C,2017A&A...600A..64T,2019MNRAS.tmp.1612T}. Instead of continuous accretion of SMBHs, it was suggested that the SMBHs have undergone multi-accretion events with the accreting gas fed with chaotic directions \citep[e.g.,][]{2006MNRAS.373L..90K,2007ApJ...667..704V}. In this case, the angular momentum acquired by the SMBH in different accretion phases cancels out in some extent, which leads to most SMBHs spinning at low or moderate rates, and then a lower average radiation efficiency of accreting SMBHs \citep*[e.g.,][]{2008ApJ...684..822B,2013IAUS..290..259L}. This may alleviate the inconsistence with the observations in some extent.

It seems no doubt that SMBHs grow dominantly through accretion of circumnuclear gas in the host galaxies \citep*[][]{1982MNRAS.200..115S,2002MNRAS.335..965Y, 2004MNRAS.351..169M,2009ApJ...690...20S, 2012MNRAS.419.2529R,2012ApJ...761....5Z}, however, there is still lack of observations constraining the origin and properties of the seed BHs, from which the SMBHs grow. Stellar mass seed BHs are naturally expected to be formed as the remnants of the Pop III stars at $z\sim 20-50$ with mass of $\sim 10-100M_\odot$ \citep*[][]{2008Sci...321..669Y,2011Sci...334.1250H, 2014ApJ...781...60H}. This kind of stellar mass BHs have been well observed in X-ray binaries in the local Universe. The growth of SMBHs from such small seed BHs through accretion needs a very long period of time \citep[][]{2016MNRAS.459.3738I,2017ApJ...850L..42P}.

Several ways were suggested to form heavy seed BHs with $\sim 10^3-10^5 M_\odot $ in the early Universe, which includes direct collapse of cloudy fragmentation into a single seed BH \citep[][]{2006MNRAS.371.1813L,2006MNRAS.370..289B,2016MNRAS.458..233L}, stellar-dynamical processes of runaway-collisions within protostellar clusters \citep[][]{2015MNRAS.451.2352K,2017MNRAS.472.1677S,2018MNRAS.476..366B}, or collapse of super-massive stars (SMS) \citep[][]{2013ApJ...774..149C,2013ApJ...778..178H,2018MNRAS.474.2757H}. However, formation of massive seeds commonly needs special environment with some strong restrictions, which implies this kind of massive seeds should not be dominant in the seed BH population \cite[][]{2008ApJ...686..801O}.

It is a well accepted scenario that a fraction of the gas at the disk surface can be accelerated into outflows by the co-rotating large scale magnetic field threading an accretion disk \citep*[][]{1982MNRAS.199..883B}. Such large scale field may probably be formed by the advection of external weak field in the accretion disk. However, it was found that the field advection in a thin turbulent accretion disk is quite inefficient due to magnetic diffusion in the disk \citep*[][]{1994MNRAS.267..235L}. Some different mechanisms were proposed for the field advection in a thin accretion disk, which include field advected by the hot corona above the disk or in the form of patches    \citep*[e.g.,][]{2005ApJ...629..960S,2009ApJ...701..885L,2009ApJ...707..428B,2012MNRAS.424.2097G}. These mechanisms can indeed alleviate the difficulty of field advection in the thin disk to some extent. \citet{2013ApJ...765..149C} suggested an alternative mechanism that most of the angular momentum and the kinetic energy of the disk can be removed by the magnetic outflows, and therefore its radial velocity is substantially increased, which leads to efficient field advection in the disk. In this model, only a fraction of the gravitational energy released in the disk is radiated out, which implies that the mass accretion rate of the disk with magnetic outflows can be much higher than the normal viscously driven disk with same luminosity \citep[][]{2013ApJ...765..149C,2016ApJ...817...71C,2016ApJ...833...30C,2019ApJ...872..149L}. Numerical calculations of the global structure of a thin disk with magnetic outflows show that the mass loss rate in the outflows is usually lower than (or comparable with) the rate of mass accreted by the BH \citep*[][]{2019ApJ...872..149L}. We conjecture that such an accretion disk predominantly driven by magnetic outflows may help rapid growth of BH in the early Universe.

In this paper, we explore the rapid growth of SMBHs at high redshifts by adopting the model of an accretion disk with magnetic outflows developed in the previous works \citep[][]{2013ApJ...765..149C,2019ApJ...872..149L}.
In our calculations, we consider the cases of continuous accretion and chaotic accretion respectively. We describe the  model in Section \ref{model}, the results of the model calculations are given in Section \ref{results}, and the last two sections contain the discussion and conclusions.

\section{Model}\label{model}

Unlike a standard thin accretion disk, most angular momentum of the gas in the disk can be centrifugally accelerated into outflows by the large scale magnetic field threading the disk, which leads to a significantly higher radial velocity of the disk than that of a normal viscous disk \citep*[][]{2013ApJ...765..149C}. As most of the gravitational energy released in the disk is carried away by the outflows, only a fraction of the energy is radiated out from the disk, which indicates that the mass accretion rate of the disk with magnetic outflows is substantially higher than that of a viscously driven normal disk with the same luminosity \citep*[][]{2016ApJ...817...71C,2016ApJ...833...30C}. If the BH at the high redshift is fed by such an accretion disk with magnetic outflows, its accretion rate can be significantly higher than the Eddington value while the disk luminosity still remains sub-Eddington. The growth of BHs in the early Universe via accretion of gas can be calculated by incorporating the disk model with magnetic outflows, in which the accretion can be either continuous or chaotic. In this case, a small seed BH with $\sim 10M_\odot$ at redshift $z\sim 20-30$ may grow to several billion solar masses at $z\sim 7$ via accretion with sub-Eddington luminosity. The model calculations can be compared with the observations.

\subsection{Growth of BH through accretion}\label{growth_bh}

The BH acquires mass as well as angular momentum through accretion. The growth of BH mass and evolution of BH spin through disk are described by
\begin{equation}\label{evolution_1}
\begin{aligned}
c^{2} \frac{d M}{d t}=&E_{\rm ms} \dot{M}_{\rm BH} \\
\frac{d J_{\rm BH}}{d t}=&J_{\rm ms} \dot{M}_{\rm BH}
\end{aligned}
\end{equation}
where $ J_{\rm ms} $ and $ E_{\rm ms} $ are the specific angular momentum and energy of the accreting matter capture by the hole at the radius of the marginal stable circular orbit $ R_{\rm ms} $ respectively \citep*[][]{1996MNRAS.283..854M}. These two quantities are given by \citep*[][]{1968PhRv..174.1559C, 1972ApJ...178..347B},
\begin{equation}\label{E_msJ_ms}
\begin{aligned}[]
E_{\rm ms}=&\frac{R_{\rm ms}^{2}-2 M R_{\rm ms} \pm a \sqrt{M R_{\rm ms}}}{R_{\rm ms}\left(R_{\rm ms}^{2}-3 M R_{\rm ms} \pm 2 a \sqrt{M R_{\rm ms}} \; \right)^{1 / 2}},\\
J_{\rm ms}=&\pm \frac{\sqrt{M R_{\rm ms}}\left(R_{\rm ms}^{2} \mp 2 a \sqrt{M R_{\rm ms}}+a^{2}\right)}{R_{\rm ms}\left(R_{\rm ms}^{2}-3 M R_{\rm ms} \pm 2 a \sqrt{M R_{\rm ms}} \;\right)^{1 / 2}},
\end{aligned}
\end{equation}
where the radius of the marginally stable orbit $ R_{\rm ms} $ is
\begin{equation}\label{}
\begin{aligned}[]
R_{\rm ms}=&\frac{R_{\rm s}}{2}\left\{3+Z_{2} \mp \left[\left(3-Z_{1}\right)\left(3+Z_{1}+2 Z_{2}\right)\right]^{1 / 2}\right\},\\
Z_{1} \equiv & 1+ \left(1-a_* ^2\right) ^{1 / 3}\left[\left(1+a_* \right)^{1 / 3}+\left(1-a_*\right)^{1 / 3}\right],\\
Z_{2} \equiv & \left(3 a_*^{2} + Z_1 ^2\right)^{1 / 2}.
\end{aligned}
\end{equation}
Using Equations (\ref{E_msJ_ms}) and the parameters in dimensionless form, we rewrite Equation (\ref{evolution_1}) as
\begin{equation}\label{evolution_2}
\begin{aligned}[]
\frac{d M}{d t}=&\tilde{E}_{\rm ms} \dot{M}_{\rm BH}, \\
\frac{d a_*}{d t}=&\frac{2\dot{M}_{\rm BH}}{M} \left[\tilde{J}_{\rm ms} - a_*\left(1 - \eta_{\rm rad}\right)\right],
\end{aligned}
\end{equation}
where the radiation efficiency of the disk is related to the specific energy of the matter at radius $ R_{\rm ms} $ by
\begin{equation}
 \eta_{\rm rad} = 1 - \tilde{E}_{\rm ms},\label{eta_rad}
 \end{equation}
for a standard thin disk, and the dimensionless parameters are defined as
\begin{equation}
\tilde{E}_{\rm ms} = \frac{E_{\rm ms}}{c^2}, \qquad \tilde{J}_{\rm ms} = \frac{J_{\rm ms}}{R_{\rm s}c}, \qquad a_* =\frac{c}{GM}J_{\rm BH},  \qquad {\rm and}  \qquad  R_{\rm s}={\frac {2GM}{c^2}}.
\end{equation}

\subsection{An accretion disk with magnetic outflows}\label{disk_outflow}

Advection of the external weak field of gas in a thin accretion disk with magnetic outflows has been well explored in \citet{2013ApJ...765..149C}, and the global structure of the accretion disk and the outflows are given in the work of \citet{2019ApJ...872..149L}, which shows that only a fraction of the gas goes into outflows compared with the gas accreted by the BH. It is found that the analytic calculations in \citet{2013ApJ...765..149C} can describe the basic features of global structure of the disk and outflows \citep*[][]{2019ApJ...872..149L}. Therefore, we mainly adopt the analytic disk model with magnetic outflows in our present work. We summarize the disk-outflow model as follows \citep*[see][for the details]{2013ApJ...765..149C}.

In cylindrical coordinates, the mass accretion rate $ \dot{M}$ reads
\begin{equation}\label{dot_M}
\dot{M} = -2\pi R\Sigma v_{R},
\end{equation}
for an accretion disk, where $\Sigma \simeq 2\rho H$ is the disk surface density, $ \rho $, $ H $ are the mean
disk density and the scale height of the disk respectively. The mass accretion rate $\dot{M}$ remains constant along radius for a disk without outflows, while it is a function of radius as part of gas is driven from the disk surface into outflows. The radial velocity of a conventional viscous accretion disk is
\begin{equation}\label{v_r_vis}
v_{R, \rm vis}=-\frac{3\nu}{2R},
\end{equation}
where the viscosity $\nu = \alpha c_s H$ is adopted, and the viscous dissipation rate per unit area of the disk surface is
\begin{equation}\label{Q_plus}
Q^{+}_{\rm vis}=\frac{1}{2} \nu \Sigma\left(R \frac{d \Omega}{d R}\right)^{2}.
\end{equation}
Substituting Equation (\ref{v_r_vis}) into (\ref{Q_plus}), we have
\begin{equation}
Q^+_{\rm vis}=-{1\over 3}R\Sigma v_{R,\rm vis}\left(R \frac{d \Omega}{d R}\right)^{2}={\frac 1{6\pi}}\dot{M}_0\left(R \frac{d \Omega}{d R}\right)^{2}, \label{Q_plus2}
\end{equation}
where Equation (\ref{dot_M}) is used.

For a thin accretion disk with strong magnetic outflows, most angular momentum of the disk is removed by the outflows, and therefore the mass accretion is predominantly driven by the magnetic torque exerted by the outflows \citep*[see][for the details]{2013ApJ...765..149C}.

The angular momentum equation of such an accretion disk with magnetic outflows is
\begin{equation}\label{angular_momentum}
 \frac{d}{d R} \left(2 \pi R \Sigma v_{R} R^{2} \Omega\right) =\frac{d}{d R}\left(2 \pi R \nu \Sigma R^{2} \frac{d \Omega}{d R}\right)+2 \pi R T_{\mathrm{m}} ,
\end{equation}
of which the first term on the right-hand side is caused by turbulent viscosity, and the second term is due to the magnetic outflows. The radial velocity $v_{R}$ of the disk derived with Equation (\ref{angular_momentum}) is
\begin{equation}\label{v_r}
v _{ R}
=-\frac{3\nu}{2R} - \frac{T_{\rm m}}{\Sigma}\left[ \frac{\partial}{\partial R}\left( R^2 \Omega\right)  \right] ^{-1}
=-\frac{3\nu}{2R} - \frac{2T_{\rm m}}{\Sigma R \Omega}=   v_{R,  {\rm vis}} + v_{R,\rm m}=(1+f_{\rm m})v_{R,  {\rm vis}},
\end{equation}
where $ T_{\rm m} $ is the magnetic torque exerted by the outflows, $v_{R,\rm m}=-2T_{\rm m}/\Sigma{R}\Omega$, and the approximation $d \Omega /dR \simeq - 3\Omega /2R$ is adopted. The parameter $f_{\rm m}\equiv v_{R,\rm m}/v_{R,\rm vis}$ describes the relative importance of the angular momentum transfer mechanisms of the disk. Substitute Equation (\ref{v_r}) into (\ref{Q_plus2}), we can derive the viscous dissipation rate of a disk with outflows as
\begin{equation}
Q^+_{\rm vis}=-{1\over 3}R\Sigma v_{R}(1+f_{\rm m})^{-1}\left(R \frac{d \Omega}{d R}\right)^{2}={\frac 1{6\pi}}\dot{M}(1+f_{\rm m})^{-1}\left(R \frac{d \Omega}{d R}\right)^{2}, \label{Q_plus3}
\end{equation}
and the luminosity of the disk is
\begin{equation}
L=\int 4\pi R Q^+_{\rm vis}dR={\frac 2{3}}\int\dot{M}(1+f_{\rm m})^{-1}\left(R \frac{d \Omega}{d R}\right)^{2}RdR. \label{lum_d}
\end{equation}
We note that the mass accretion rate $\dot{M}$ varies with radius in the disk with magnetic outflows. As most radiation flux being radiated within several ten Schwarzschild radii of the BH, the disk radiation is predominately dependent on the mass accretion rate at the inner region of the disk. It is found that only a small fraction of gas in the inner region of the disk is driven into the outflows \citep*[see][for the details]{2019ApJ...872..149L}, and therefore the mass accretion rate at the inner region of the disk is very close to the rate accreted by the BH. The disk luminosity can be approximated as
\begin{equation}
L\approx {\frac 2{3}}\dot{M}_{\rm BH}\int(1+f_{\rm m})^{-1}\left(R \frac{d \Omega}{d R}\right)^{2}RdR, \label{lum_d2}
\end{equation}
where $\dot{M}_{\rm BH}$ is the rate of mass accreted by the BH. Comparing it with the luminosity of a viscous disk without outflows,
\begin{equation}
L_0={\frac 2{3}}\dot{M}_{0}\int\left(R \frac{d \Omega}{d R}\right)^{2}RdR, \label{lum_d0}
\end{equation}
we have
\begin{equation}
L\approx (1+\bar{f}_{\rm m})^{-1}L_0,
\label{lum_d3}
\end{equation}
where
\begin{equation}
(1+\bar{f}_{\rm m})^{-1}\approx\left. \int(1+f_{\rm m})^{-1}\left(R \frac{d \Omega}{d R}\right)^{2}RdR\middle/\int\left(R \frac{d \Omega}{d R}\right)^{2}RdR\right..
\label{bar_f_m}
\end{equation}
In principle, $f_{\rm m}\equiv v_{R,\rm m}/v_{R,\rm vis}$ is a function of radius. However, in the present accretion disk model with magnetic outflows, $f_{\rm m}$ remains nearly constant radially in the inner region of the disk, i.e., the radial velocities of the disks with outflows are nearly in parallel with that of a viscous disk without outflows in Figure 2 of \citet{2019ApJ...872..149L}, therefore we have $\bar{f}_{\rm m}\approx f_{\rm m}$. If $\dot{M}_{\rm BH}=\dot{M}_0$, which means that the luminosity of the disk with outflows is about $(1+f_{\rm m})^{-1}$ times that of a viscous accretion disk without outflows if the gas is accreted at the same rate by the BH, or
\begin{equation}
\dot{M}_{\rm BH}\approx (1+f_{\rm m})\dot{M}_0,
\label{mdot}
\end{equation}
if $L=L_0$, i.e., the mass accretion rate of the BH surrounded by a disk with outflows should be about $(1+f_{\rm m})$ times that of a viscous accretion disk in order to radiate at the same luminosity. This is caused by the fact that part of gravitational energy released in the disk is tapped into the outflows.

It is well known that the acceleration of the outflows by the field co-rotating with the disk is sensitive to the field line inclination at the disk surface. For cold gas driven by the magnetic field from a Keplerian rotating thin disk, the inclination angle of the field line with respect to the disk plane requires $\theta\le 60 ^ {\degr}$ \citep*[][]{1982MNRAS.199..883B}. In more realistic cases, for example, considering the gas pressure or/and slightly sub-Keplerian motion of the disk, the critical field inclination required for launching outflows does not deviate much from $60^\circ$ \citep*[][]{1994A&A...287...80C,1998ApJ...499..329O,2014ApJ...783...51C}. The magnetic field dragged by the accretion disk is described by the induction equation \citep*[e.g., see Equation 14 in][]{1994MNRAS.267..235L}. The field inclination at the disk surface can be estimated by equating the advection timescale to the diffusion timescale of the field for a steady disk, which are
\begin{equation}
\tau_{\rm adv} \sim -\frac{R}{v_R}, \qquad {\rm and} \qquad \tau _{\rm dif} \sim \frac{RH \kappa_0}{\eta}, \label{tau}
\end{equation}
respectively, where $\kappa_0=B_z/B_R^{\rm s}$ at the disk surface, and $\eta$ is the magnetic diffusivity \citep*[see][for the details]{1994MNRAS.267..235L}. Substitute Equations (\ref{v_r_vis}) and (\ref{v_r}) into Equation (\ref{tau}) and let $\tau_{\rm adv}=\tau _{\rm dif}$, we obtain
\begin{equation}
1+f_{\rm m}={\frac {2}{3}}{\cal{P}}_{\rm m}\kappa_0^{-1}\left({\frac {H}{R}}\right)^{-1}. \label{f_m}
\end{equation}
where the magnetic Prandtl number $ {\cal{P}}_{\rm m} = \eta/\nu $. It was found that the magnetic Prandtl number $ {\cal{P}}_{\rm m}$ is always around unity for isotropic turbulent plasma either estimated by order of magnitude or numerical simulations \citep*[e.g.,][]{Parker..1979,2003A&A...411..321Y,2009A&A...507...19F,2009ApJ...697.1901G}. The field inclination $\kappa_0$ at the surface of the inner disk changes very little radially \citep*[see Figure 1 in][]{2019ApJ...872..149L}, which leads to roughly constant $f_{\rm m}$ in the inner disk region (see Equation \ref{f_m}). We find that $f_{\rm m}\sim 5$ if typical values of parameters, ${\cal{P}}_{\rm m}=1$, $H/R\sim0.1$, and $\kappa_0\sim 1$ are adopted.

\section{results}\label{results}

%fig 1

\begin{figure}
	\centering
	\includegraphics[width=0.5\columnwidth]{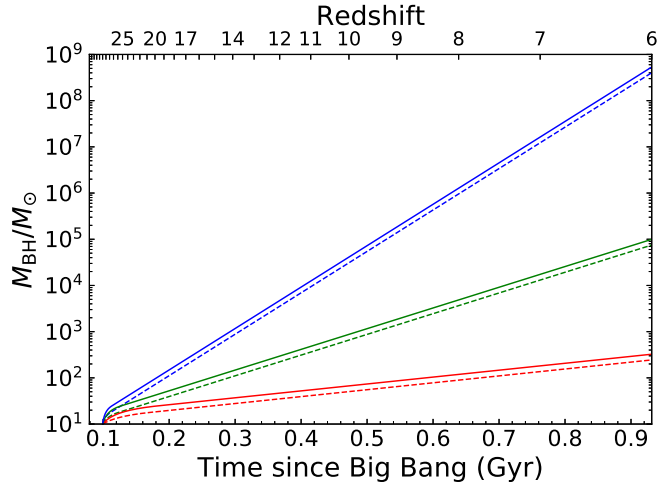}
	\caption{BH mass evolution with the cosmic time via an accretion disk radiating at the Eddington limit, $ \lambda = 1$. The solid lines correspond to the evolution starting from $a_{*,\rm init}=0$, while the dashed lines show the evolution starting from $a_{*,\rm init}=0.75$. The seed BH is assumed to be formed as the remnant of Pop III stars at redshift $ z\sim 30 $  with $ M_{\rm seed}=10M_{\odot} $. The red lines correspond to the accretion driven by turbulent viscosity (i.e., without outflows $ f_{\rm m}=0$), while the green lines ($ f_{\rm m}=2$) and blue lines ($ f_{\rm m}=5$) indicate the accretion disk driven by both of the turbulent viscosity and the magnetic outflows, $f_{\rm m}$ is defined in Equation (\ref{v_r}).}
	\label{BH_mass_Edd}
\end{figure}

%fig 2

\begin{figure}
	\centering
	\includegraphics[width=0.5\columnwidth]{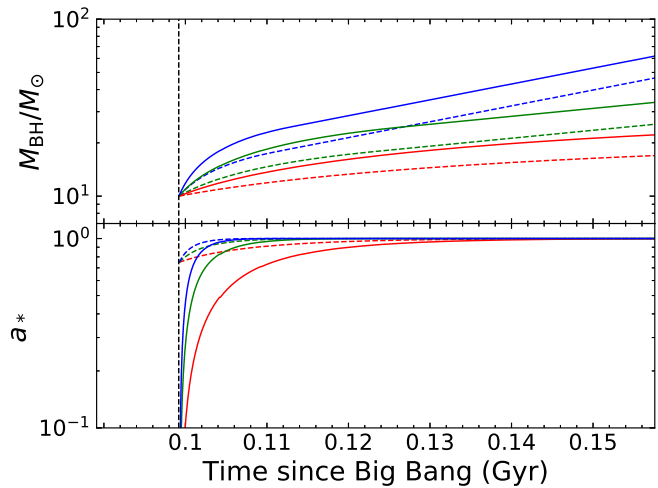}
	\caption{The same as Figure \ref{BH_mass_Edd}, but also plot for the evolution of spin parameter $ a_* $ with cosmic time, the vertical dashed line corresponds to the redshift at $ z=30 $.  }
	\label{BH_a_Edd}
\end{figure}

%fig 3

\begin{figure}
	\centering
	\includegraphics[width=0.5\columnwidth]{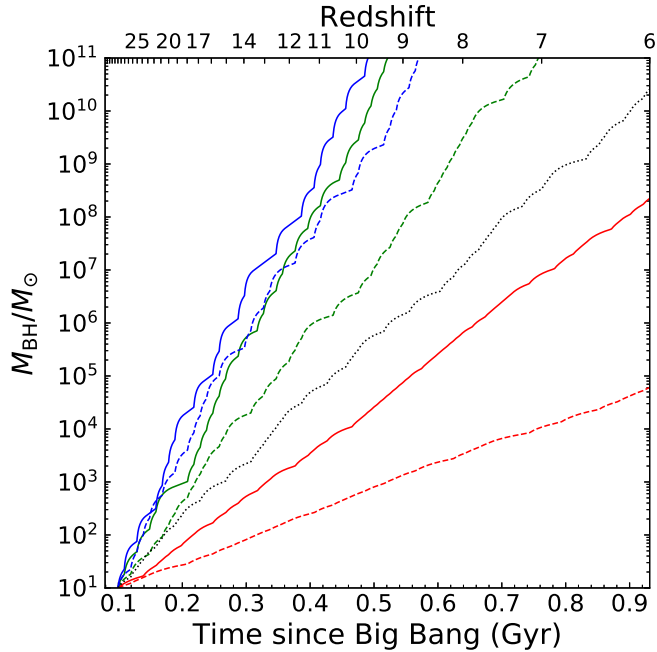}
	\caption{BH mass evolution along the cosmic time via chaotic accretion, in which the duration of each accretion episode $ 10^7 \rm yr $ is adopted. The solid and dashed lines correspond to different outflows, $ f_{\rm m}=2 $ (dashed), and $ 5 $ (solid), respectively. The color lines show the results for different Eddington ratios, $ \lambda=0.1 $ (red), $ 0.5 $ (green), and $ 1$ (blue), respectively. The dotted line shows the accretion disk radiating at Eddington luminosity without outflows (i.e., $ f_{\rm m}=0 $).  The initial value of the spin parameter of the seed BH is set to $ a_{*, \rm init}=0 $, and the mass of the seed BH $ M_{\rm seed}=10M_{\odot} $ is adopted.}
	\label{BH_mass_1_10_7}
\end{figure}

%fig 4

\begin{figure}
	\centering
	\includegraphics[width=0.9\columnwidth]{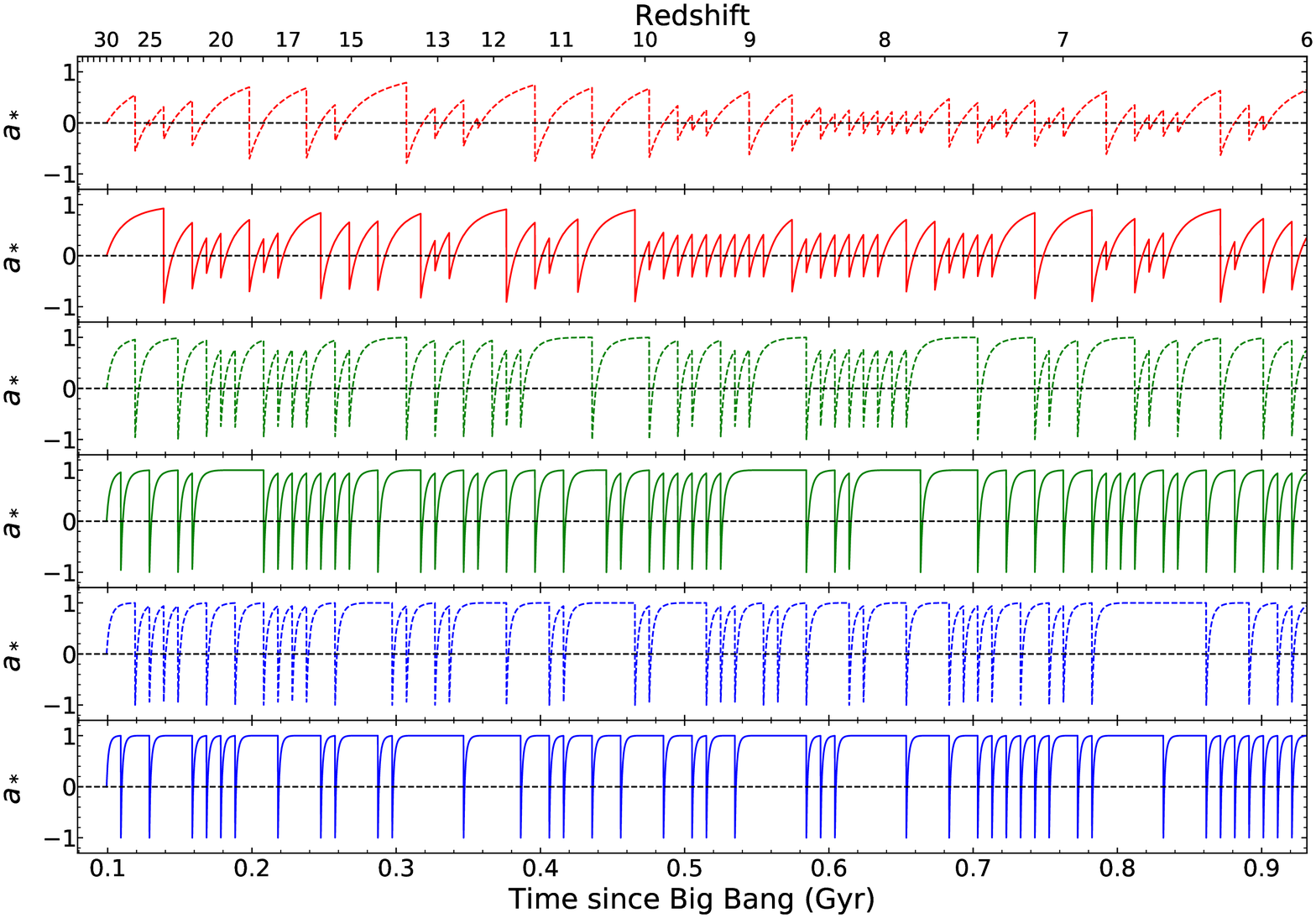}
	\caption{BH spin evolution along the cosmic time via chaotic accretion, the duration of each accretion episode $ 10^7 \rm yr $ is adopted. The solid and dashed lines correspond to different outflows, $f_{\rm m}=2 $ (dashed), $ 5 $ (solid). The color lines show different Eddington ratios, $ \lambda=0.1 $ (red), $ \lambda=0.5 $ (green), and $ \lambda=1 $ (blue). The initial value of the spin parameter of the seed BH is set to $ a_{*, \rm init}=0 $, and its mass $ M_{\rm seed}=10M_{\odot} $ is adopted. Here, note that for a prograde rotating BH $a_{*}>0$ and a retrograde rotating BH $a_{*}<0$.}
	\label{BH_a_1_10_7}
\end{figure}

%fig 5

\begin{figure}
	\centering
	\includegraphics[width=0.5\columnwidth]{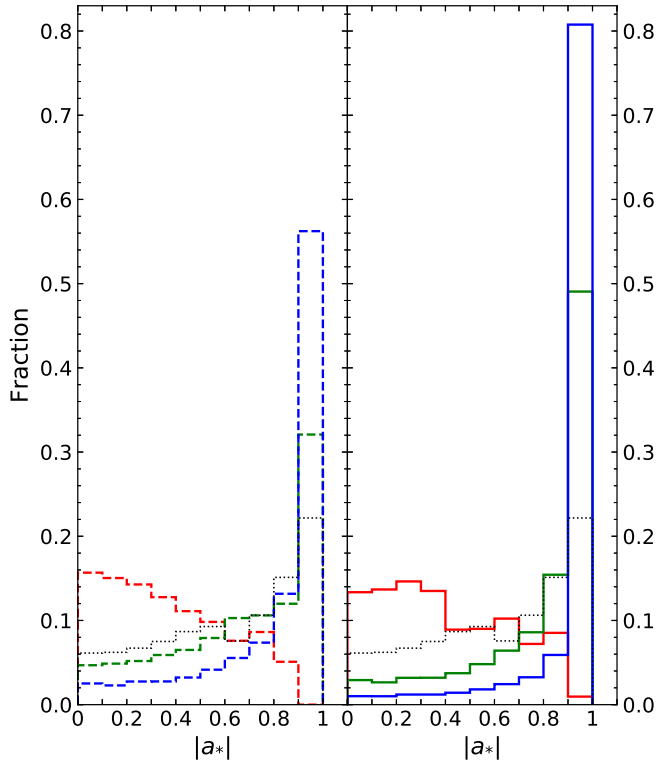}
	\caption{Distributions of BH spin parameter during the time of the BH grows (i.e., only the BH mass $ M_{\rm BH} < 10^{11}M_{\odot}$). The solid and dashed lines correspond to different magnetic outflows with $ f_{\rm m}=2 $ (dashed) and $ 5 $ (solid), respectively. The color lines show different Eddington ratios, $ \lambda=0.1 $ (red), $ 0.5 $ (green), and $ 1$ (blue), respectively. The dotted line shows the accretion disk without outflows (i.e., $ f_{\rm m}=0 $).  The initial value of the spin parameter of the seed BH is set to $ a_{*, \rm init}=0 $. The mass of the seed BH $ M_{\rm seed}=10M_{\odot} $ and  the duration of each accretion episode $ 10^7 \rm yr $ are adopted.}
	\label{BH_Stat_a_1_10_7}
\end{figure}

%fig 6

\begin{figure}
	\centering
	\includegraphics[width=0.5\columnwidth]{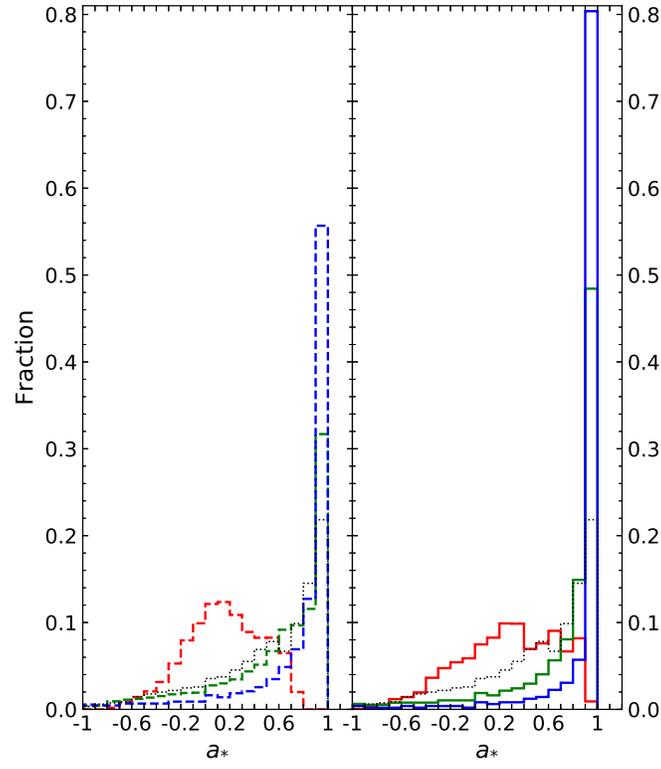}
	\caption{The same as Fig. \ref{BH_Stat_a_1_10_7}, but the distributions of the spin parameter are counted from $ a_{*}=-1 $ to $ 1 $ during the BH growth.}
	\label{BH_Stat_a_1_10_7_1}
\end{figure}

%fig 7

\begin{figure}
	\centering
	\includegraphics[width=0.5\columnwidth]{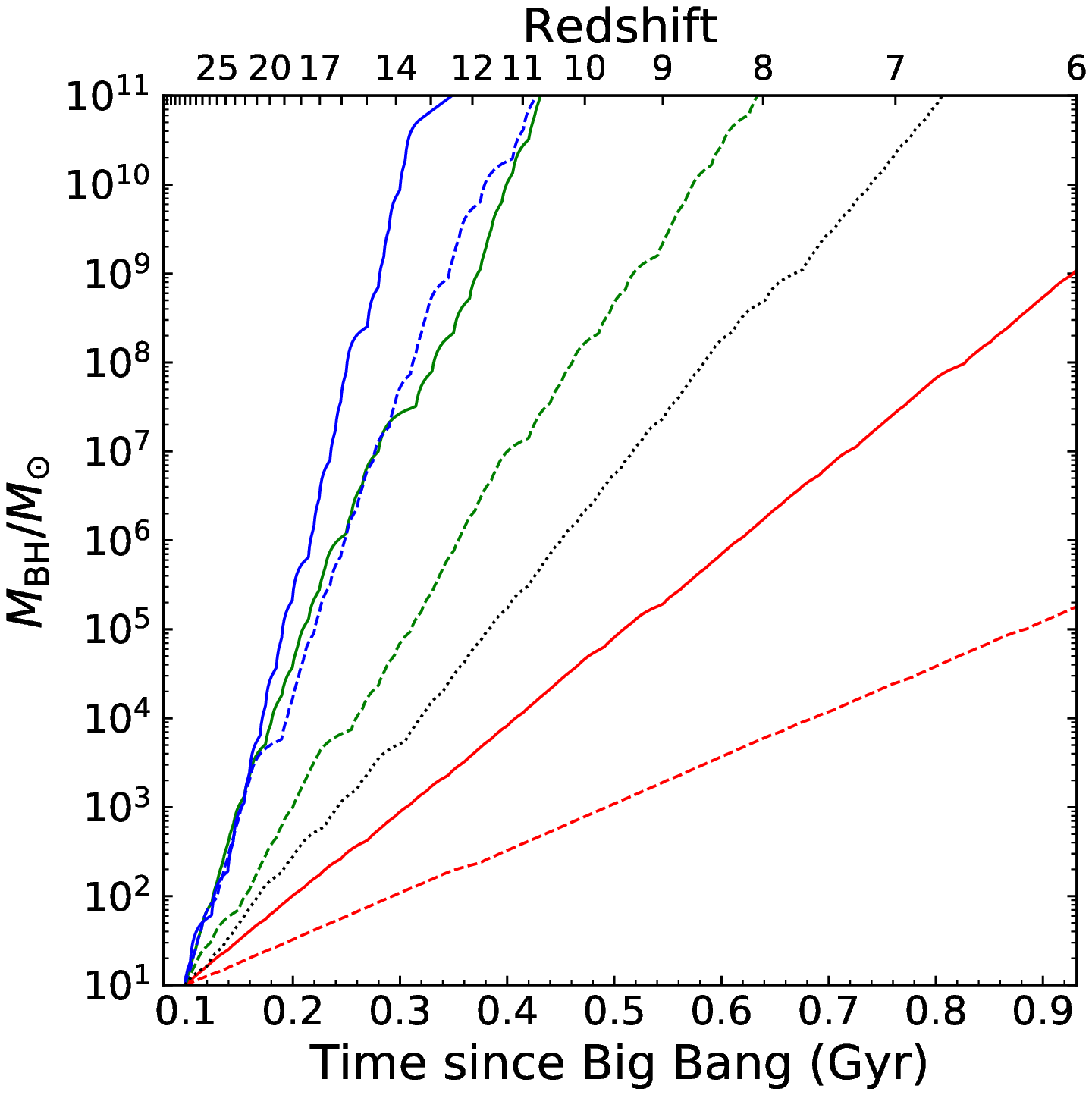}
	\caption{The same as Figure \ref{BH_mass_1_10_7}, but the duration of each accretion episode $ 5 \times 10^6 \rm yr $ is adopted. }
	\label{BH_mass_5_10_6}
\end{figure}

The realistic SMBH growth must be very complicated, on which one has to consider accretion of gas and mergers of BHs  \citep*[e.g.,][]{2001ApJ...558..535M,2003MNRAS.340..647I,2003ApJ...582..559V,2004ApJ...614L..25Y,2005ApJ...620...59S,2009ApJ...696.1798T}. In this work, we focus on the BH growth in early Universe through an accretion disk with magnetic outflows. The global structure of an accretion disk with magnetically driven outflows is extensively explored by \citet{2019ApJ...872..149L}. To avoid the complexity of the calculations, we adopt the quantity $f_{\rm m}$, the relative importance of angular momentum transfer rate by the outflows, as an input model parameter, the value of which is well constrained either by the numerical results in \citet{2019ApJ...872..149L} or the estimate given in Section \ref{disk_outflow}. The seed BH is crucial for the growth of SMBHs at high redshifts. In this work, we conservatively assume that the seed BHs with $M_{\rm seed}=10M_\odot$ are formed as the remnants of Pop III stars at redshift $z=30$. Our present calculations are also applicable for more massive seed BHs, and any heavier seed BHs are definitely more helpful for the growth of SMBHs at high redshifts.

The disk accreting at a high rate with super-Eddington luminosity is regarded as an efficient way to feed SMBHs at high redshifts. With a super-Eddington accretion rate, however, most of the gas in the disk may be blown away into outflows and/or jets, and the BH would swallow the gas at a rate probably not more than a few times Eddington rate \citep[\citealt{2016MNRAS.459.3738I}, and see][for a review]{2013ASSL..396..293H}. The photon trapping effect become efficient when the BH accreting at such a high rate, most of the released energy is trapped in the gas and goes accreted by the BH, thus the radiation efficiency is only about a few percent \citep[][]{2014ApJ...796..106J,2017arXiv170902845J}, which is smaller than the average efficiency $\sim 0.09 - 0.2$ suggested by the numerical calculations/simulations with the constraints of observations \cite[see][for details and references therein]{1982MNRAS.200..115S,2002MNRAS.335..965Y, 2004MNRAS.351..169M, 2009ApJ...690...20S, 2012MNRAS.419.2529R,2012ApJ...761....5Z}.  Thus, we assume that the luminosity of the accretion disk with outflows is Eddington-limited in this work, i.e., the parameter $\lambda=L/L_{\rm Edd}\le 1$ [$L_{\rm Edd}\equiv 1.3\times 10^{38}(M_{\rm BH}/M_\odot)~ {\rm erg~s^{-1}}$] is adopted.

In the case of a BH with mass $M_{\rm BH}$ spinning at $a_*$, the mass accretion rate $\dot{M}_{\rm BH}$ of the disk with magnetic outflows can be derived with Equations (\ref{eta_rad}) and (\ref{mdot}) provided the values of parameters $\lambda$ and $f_{\rm m}$ are specified. The evolution of the BH mass and spin is then calculated with Equation (\ref{evolution_2}). In Figure \ref{BH_mass_Edd}, we show how the seed BH grows with time through continuous  accretion of gas near the BH with Eddington luminosity. It is found that the seed BH with $10M_\odot$ can only grow to a few hundred solar mass BH at $z\sim 6$ if the gas is accreted in a normal viscous disk. In the case of an accretion disk with magnetic outflows, the BH mass can be as high as several million solar mass at $z\sim 6$ for strong magnetic outflows (e.g., $f_{\rm m}=5$). The evolution of the BH spin parameter varying with time is plotted in Figure \ref{BH_a_Edd}. It is found that the BH will become an extreme Kerr BH within a very short period of time, and therefore the radiation efficiency reaches its maximal value correspondingly. As the accretion luminosity is assumed to be Eddington-limited, a high radiation efficiency means a lower mass accretion rate, which leads to a slower growth of BH.

It was suggested that the growth of BH at the center of galaxies may experience multi-accretion events. The angular momentum of the gas feeds the BH at random directions with respect to the BH spin axis in each episode accretion \citep*[][]{2006MNRAS.373L..90K,2007ApJ...667..704V}. For simplicity, we only consider the cases of the disk axis is aligned with the BH spin, namely, the gas of the disk is either in direct orbits or retrograde orbits for each episode accretion event. We adopt the duration of each accretion episode as an input parameter. In Figure \ref{BH_mass_1_10_7}, we plot the growth of BHs through episode accretion with different modes, i.e., a normal viscous disk or a disk with magnetic outflows. The duration of each episode accretion, $10^7$~years, is adopted in the calculations. It is found that the seed BH is not able to grow to a billion solar mass BH at $z\sim 7$ through a viscous disk without magnetic outflows, while the BH grows very rapidly to a SMBH through accretion in a disk with magnetic outflows even if the disk is radiating at a sub-Eddington luminosity.

In Figure \ref{BH_a_1_10_7}, we show the time evolution of BH spin parameter $a_*$ during the growth of BH through accretion calculated with different values of model parameters, from which the distributions of the BH spin parameter $a_*$ can be derived (see Figures \ref{BH_Stat_a_1_10_7} and \ref{BH_Stat_a_1_10_7_1}). Only the BHs with $M<10^{11}M_\odot$ are counted in the calculations of the distributions in the early Universe with $z\ge 6$. It is found that most BHs are spinning rapidly when $\lambda=1$ is adopted for the accretion disks with outflows, while the values of spinning parameter $a_*$ are nearly homogeneously distributed in the low-$\lambda$ cases. The results calculated with the episode duration of $5\times 10^6$~yr are given in Figure \ref{BH_mass_5_10_6}.

It was claimed that, the gas in the accretion disk plunges onto the black hole at the radius
around the marginal stable circular orbit, which means that the magnetic field diffusion is negligible in this plunging region, and the field flux may be substantially enhanced in this region \citep*[][]{2006ApJ...651.1023R,2009ApJ...699..400G}. It implies that the jets produced through the Blandford-Znajek mechanism may be more powerful for a retrograde rotating BH than a prograde counterpart \citep*[][]{2006ApJ...651.1023R,2009ApJ...699..400G}. However, this seems to contradict the results of the numerical simulations carried out by \citet{2012MNRAS.423L..55T}, in which the radial velocity of the geometrically thick disk is found to be sufficiently large to capture magnetic fluxes in the region near the BH. Although it is still a debating issue, we plot the distributions of the spin parameter in the range of $-1<a_*<1$ in Figure \ref{BH_Stat_a_1_10_7_1}. We summarize the results of spin parameter distributions in Table \ref{table1}. We find that the fraction of the BHs with high-$a_*$ is lower than the case with episode duration of $10^7$~yr, if the values of all other model parameters are fixed.

%==================

\begin{table*}[h]
	
	\centering
	\caption{Distributions of BH spin parameter during the time of the BH growth. \label{table1}}
	\begin{tabular}{ccccccc}
		\toprule
		\toprule
		\multirow{2}{*}{\tabincell{c}{Episodic accretion duration\\(yr)}} &
		\multirow{2}{*}{$\lambda$} &
		\multirow{2}{*}{$f_{\rm m}$} &
		\multicolumn{4}{c}{Fraction}  \\
		\cmidrule{4-7}
		& & &|$a_*|\leqslant0.1$ & $0.1<|a_*|\leqslant0.5$ &$0.5<|a_*|\leqslant 0.9$&$|a_*|>0.9$\\
		\midrule
		\multirow{7}{*}{$10^7$}&
		\multirow{2}{*}{$0.1$} &
		2&0.157&0.532&0.311&0.000\\
		&&5&0.134&0.508&0.349&0.009\\
		\cmidrule{2-7}
		&\multirow{2}{*}{$0.5$} &
		2&0.047&0.226&0.407&0.320\\
		&&5&0.029&0.128&0.352&0.491\\
		\cmidrule{2-7}
		&\multirow{3}{*}{$1$} &
		0&0.062&0.292&0.425&0.221\\
		&&2&0.026&0.110&0.302&0.562\\
		&&5&0.011&0.048&0.134&0.807\\
		\midrule
		\multirow{7}{*}{$5\times 10^6$}&
		\multirow{2}{*}{$0.1$} &
		2&0.336&0.615&0.049&0.000\\
		&&5&0.212&0.619&0.169&0.000\\
		\cmidrule{2-7}
		&\multirow{2}{*}{$0.5$} &
		2&0.083&0.373&0.394&0.150\\
		&&5&0.047&0.210&0.425&0.318\\
		\cmidrule{2-7}
		&\multirow{3}{*}{$1$} &
		0&0.126&0.445&0.395&0.034\\
		&&2&0.052&0.224&0.408&0.316\\
		&&5&0.023&0.102&0.278&0.597\\
		\bottomrule	
	\end{tabular}
	\tablecomments{$\lambda=L/L_{\rm Edd}$ is the Eddington ratio, $a_*$ is the dimensionless spin parameter, and $f_{\rm m}$ is defined in Equation (\ref{v_r}).}
\end{table*}

%==========================

\section{discussion}\label{conclusion}

The motivation of this work is to explore how SMBHs grow from stellar mass seed BHs at high redshifts. It was suggested that the mass accretion rate of a slim disk can be higher than Eddington rate, which radiates at super-Eddington luminosity. However, such a slim disk with super-Eddington luminosity will inevitably accelerating the gas at the disk surface into outflows, and only a small fraction of the gas accreted in the disk is ultimately swallowed by the BH even if the mass accretion rate is very high at the outer edge of the disk \citep[e.g.,][]{1999MNRAS.310.1002S,2002ApJ...573..738H,2003ApJ...592.1042I,2014ApJ...780...79Y,2015MNRAS.448.3514C}. \citet{2002ApJ...568L..97B} proposed that a thin disk with small scale inhomogeneities may radiate super-Eddington fluxes, which may alleviate the difficulty of rapid growth of the SMBH at high redshifts to some extent \citep*[][]{2013ASSL..396..293H}. In the calculations of this work, we find that the BH growth is rather inefficient through a normal disk with Eddington luminosity (see Figure \ref{BH_mass_Edd}), which is consistent with some previous works \citep*[e.g.,][] {2004ApJ...614L..25Y,2005ApJ...620...59S}. This implies a large seed BH with $10^{4-5} M_\odot$ is required to grow to a billion solar mass BH at $z\sim 7$ through a normal Eddington luminosity limited accretion disk \citep*[][]{2005ApJ...620...59S,2013ApJ...771..116J,2016MNRAS.457.3356V}. Such difficulty of rapid BH growth is alleviated if magnetically driven outflows are present, which may carry away most angular momentum and the gravitational power of the gas in the disk. The accretion rate of the disk is therefore significantly higher than the Eddington rate, while it radiates at sub-Eddington luminosity (see Section \ref{disk_outflow}). We find that a stellar mass seed BH at $z=30$ can be grown to $\sim 10^7 M_\odot$ at $z=7$ through continuous Eddington luminosity limited accretion predominantly driven by strong magnetic outflows (see Figure \ref{BH_mass_Edd}), which is still too small compared with the observed quasar with $\sim 10^9 M_\odot$ at $z=7.54$ \citep*[][]{2018Natur.553..473B}. It is well known that the BH can be spun up very rapidly through accretion, and the radiation efficiency of the disk surrounding a rapidly spinning BH is significantly higher than that for a non-rotating BH, which leads to a lower mass accretion rate for the Eddington luminosity limited disk surrounding a spinning BH (see Figure \ref{BH_a_Edd}). This makes the BH grow slowly while it is spun up by continuous accretion even if strong magnetic outflows are present. One inevitable consequence of the BH growth through such persistent accretion is most SMBHs being spinning very rapidly \citep*[][]{2007ApJ...667..704V}. For the accretion disk with magnetically driven outflows, a strong magnetic field is present near the BH, and jets may probably be formed through the Blandford-Znajek mechanism \citep*[][]{1977MNRAS.179..433B}.  It implies that jets should appear in most accreting SMBHs at least at high redshifts, which seems to be inconsistent with the observations of only a small fraction of quasars being radio-loud \citep*[][]{1989AJ.....98.1195K}. This may be alleviated if episodic chaotic accretion is responsible for quasars instead of persistent accretion \citep*[][]{2008MNRAS.385.1621K,2013ApJ...775...94V}.

We incorporate our accretion disk model with magnetically driven outflows into the chaotic accretion scenario. It is found that a stellar mass BH at $z=30$ can grow to $\ga 10^9 M_\odot$ at $z\sim 8$ through chaotic accretion at a moderate rate ($\lambda=0.5$) with mild magnetic outflows ($f_{\rm m}=2$) (the duration of each accretion episode $10^7$~yr is adopted, see Figure \ref{BH_mass_1_10_7}). We find that most BHs are spinning at $a_*>0.9$ in the case of the disk radiating at $\lambda=1$, while the moderately spinning BHs are dominant if $\lambda\la0.5$ is adopted (see Figures \ref{BH_Stat_a_1_10_7} and \ref{BH_Stat_a_1_10_7_1}), which is caused by the fact that the BH can be spun up to a higher spin parameter $a_*$ in each accretion episode for a higher $\lambda$ adopted. There are several methods have been proposed to constrain the episodic timescale of AGNs \citep*[see][for a review]{2004cbhg.symp..169M}, though the results are quite uncertain. Typical duration of episodic accretion event around $ 10^6 \sim 10^7 \rm yr $ inferred from the observations have been adopted in some previous works (see \citealt{2015ApJ...804...45L}, Table 1.,\citealt{2002MNRAS.335..965Y}, \citealt{2002ApJ...578..702H}), which is also consistent with the calculations in \cite{2006MNRAS.373L..90K}. We also perform calculations of the BH growth through chaotic accretion with a short duration of each episode accretion being $5\times 10^6$ yr. It is found that most BHs are spinning at moderate values of $a_*$ even if a high $\lambda\sim 1$ is adopted (see Table \ref{table1}). This is because a lower final value of spin parameter after each accretion episode is caused by a shorter accretion duration. The distributions of $a_*$ derived in this work are qualitatively consistent with the numerical simulations of cosmological evolution of SMBHs in \cite{2007ApJ...667..704V}.

Most disk radiation is from the inner region of the disk with several ten Schwarzschild radii to the BH. The typical mass accretion rate in this region is very close to the rate swallowed by the BH \citep*[see][]{2019ApJ...872..149L}. Thus, the mass/spin evolution of the BH is almost independent of the mass loss rate in the outflows, and this will not alter our main conclusions of this work.  

We note that the structure of the inner edge of the disk is quite complicated \citep[see][for the details]{2000ApJ...528..161A,2002ApJ...573..754K,2010A&A...521A..15A}. Unlike a standard thin disk with no stress at its inner edge, it was suggested that a non-zero stress at the inner edge of the disk in some previous works \citep[see][for the details]{2000ApJ...528..161A,2002ApJ...573..754K}. This makes the radiation efficiency of the disk higher than that of a standard thin disk, however, the detailed physics of such disk model is still quite uncertain, which prevents us from doing further calculations based on their model. In principle, a higher radiation efficiency will make the BH growth slower, which may require stronger magnetic outflows in our model calculations to reproduce the observed luminous quasars at high redshifts. \citet{2010A&A...521A..15A} explored the inner edge of the disk at various luminosities, and they found that the disk may extend into the innermost stable circular orbit (ISCO), while the specific angular momentum of the gas in the disk region within the ISCO almost remains the Keplerian value at the ISCO \citep*[see Figure 15 in][]{2010A&A...521A..15A}. It implies that the BH evolution equations in Section \ref{growth_bh} can indeed describe the mass/spin evolution of the BH quite well.

The disk spectrum depends on it structure, and how the spectrum would be affected by the deviation of the disk structure from a standard thin disk has been explored in some previous works \citep*[e.g.,][]{2008ApJ...676..549S,2014MNRAS.438.3024L,2016ApJ...821..104Y}, in which different outflow driven mechanisms have been considered \citep*[][]{2014MNRAS.438.3024L,2016ApJ...821..104Y}. It was suggested that the line driven outflows from disks surrounding massive BHs may help reproduce the observed UV spectra in AGNs \citep*[][]{2014MNRAS.438.3024L}. In their model, a substantial fraction of mass in the disk is driven into outflows, which significantly reduces the gravitational power released in the disk, and then the disk temperature and luminosity. In their calculations, the specific angular momentum of the gas at the disk surface driven into the outflows is nearly the same as that of the gas rotating in the disk, which is quite different from the magnetically driven outflows considered in this work. For magnetically driven outflows, the gas moves along the magnetic field line co-rotating with the disk roughly approaching the Alfven radius, which is usually much larger than the radius of the field line footpoint at the disk surface \citep*[see][for the details]{spruit96}. The specific angular momentum of the ultimately accelerated gas in the outflows is therefore much higher than that when it is leaving from the disk surface. Thus, the gas in the outflows may carry away most of the angular momentum of the disk, which makes the radial velocity of the disk much higher than that of a viscous disk without outflows. The disk with magnetic outflows is accreting at a higher rate compared with a viscously driven disk with the same luminosity, because most gravitational power released in the disk is tapped into the outflows by the magnetic field co-rotating with the disk. A variety of different outflow models, including Blandford-Payne outflow model, have been employed in the disk spectral calculations \citep*[][]{2016ApJ...821..104Y}. The angular momentum of the disk carried away by the magnetic outflows has not been considered in their work due to a rather simplified magnetic outflow model is used, which may not affect their spectral calculations, because the local structure of the disk (temperature and density) is unchanged except its radial velocity, if the angular momentum of the gas is properly considered as this work. The growth of massive BHs via accretion disks with different types of outflows other than the magnetic outflows considered in this work is expected  to be investigated in the future, which is beyond the scope of this work.

\section{conclusions}
Our results show that persistent accretion with strong magnetic outflows is unable to grow a billion solar mass BH discovered at a high redshift $z\sim 7$ from a stellar mass seed BH with $10M_\odot$ at $z=30$, while such SMBHs with several billion solar masses may probably be grown by chaotic accretion predominantly driven by magnetic outflows. It is found that disks radiating at moderate luminosity ($\lambda\sim 0.5$) with mild outflows ($f_{\rm m}\sim 2$) can reproduce the observed SMBHs at high redshifts. In this work, we consider the simplest case, the accretion disk is in alignment with the BH axis (either co-rating or counter-rotating with the BH). In the realistic case, the direction of the gas motion near the BH may be randomly distributed. Due to the Lense-Thirring effect, the inner disk tends to be aligning with the equatorial plane of the BH \citep*[][]{1975ApJ...195L..65B}. On the other hand, the BH is being spun up by the accreting gas, which also decreases the misalignment between the BH spin and the accretion axis. As we are focusing on the mass growth of SMBHs, this is a fairly good approximation for our present investigation. We only consider the BH growth through accretion in this work, and it is obviously that the mergers of BHs may also play important roles in BH growth \citep*[][]{2001ApJ...558..535M,2003ApJ...582..559V,2008ApJ...684..822B,2009ApJ...696.1798T,2007Sci...316.1874M,2015ApJ...810...51M}. The calculations will be more complicated if the mergers of BHs are included, however, it is clear that mergers will help mass growth of BHs. Thus, the BH growth will be more efficient while the mergers are properly included in the calculations, which will not alter the main conclusions of this work.

\acknowledgments

We thank the referee for his/her helpful comments/suggestions. This work is supported by the NSFC grants (11773050 and 11833007), the CAS grant (QYZDJ-SSWSYS023), and a start-up grant from Zhejiang University.

\end{document}